# Smart Cane: Assistive Cane for Visually-impaired People

Mohd Helmy Abd Wahab[1], Amirul A. Talib[1], Herdawatie A. Kadir[1], Ayob Johari[1], A.Noraziah[2], Roslina M. Sidek[2], Ariffin A. Mutalib[3]

[1] Faculty of Electrical and Electronic Engineering, Universiti Tun Hussein Onn Malaysia
86400 Batu Pahat, Johor, Malaysia

[2] Faculty of Software Engineering and Computer System, Universiti Malaysia Pahang
26300, Kuantan, Pahang, Malaysia

[3] College of Arts and Sciences, Universiti Utara Malaysia
06010 Sintok, Kedah, Malaysia

## Abstract

This paper reports on a study that helps visually-impaired people to walk more confidently. The study hypothesizes that a smart cane that alerts visually-impaired people over obstacles in front could help them in walking with less accident. The aim of the paper is to address the development work of a cane that could communicate with the users through voice alert and vibration, which is named Smart Cane. The development work involves coding and physical installation. A series of tests have been carried out on the smart cane and the results are discussed. This study found that the Smart Cane functions well as intended, in alerting users about the obstacles in front.

## 1. Introduction

The increasing number of people with disabilities in Malaysia attracts the concern of researchers to invent various technologies, hoping that these technologies can assist the disabled people in carrying out their tasks in everyday life like normal people [1]. The increasing trend of people with disabilities has been reported by Country Report Malaysia, the 7th ASEAN and Japan High Level Officials Meeting on Caring Societies in 2009. As stated in the report, in 2008, there were only 30,522 children with disabilities detected. The amount increased to 13.7% in 2009 where 35,368 people with disabilities were registered with the Department of Social Welfare [2]. Therefore, technologies are considered the best solution to assist this community to perform tasks as the normal people.

The objective of the paper is to discuss the development work of a cane that could communicate with the users through voice alert and vibration, which is named Smart Cane which involves coding and physical installation.

This paper is organized as follows; section 2 describes some related work. Methodology is explained in section 3. Results presented in section 4 and the paper is concluded in section 5.

## 2. Related Work

This section describes appropriate related works on the development of smart canes intended for visually-impaired people.

According to [1], technology can help in reducing many barriers that people with disabilities face. These kinds of technologies are referred to as assistive technology (AT). There are many types of disabilities, including physical disabilities, hearing-impaired, and visually-impaired. AT has been utilized in assisting them [4, 5, 6, 9]. However, developing an AT is expensive [3], making their selling price high.

According to Mazo and Rodriguez [8] the blind Cane is one of the assisting tools for the visually-impaired and it is really important. According to Herman [3], one of the main problems of the visually-impaired, is that most of these people have lost their physical integrity. Also, they do not have confidence in themselves. This statement has been proven by Bouvrie [7], in which an experiment name "Project Prakash" has been carried out. It was intended at testing the visually-impaired to utilize their brain to identify set of objects. According to Chang and Song [12], this can also be applied to different situation. When the visually-impaired walk into a new environment, they will find it difficult to memorize the locations of the object or obstacles. These examples demonstrate the difficulties of visually-impaired people.

The GuideCane [9] is designed to help the visually-impaired users navigate safely and quickly among obstacles and other hazards. GuideCane is used like the widely used white cane, where the user holds the GuideCane in front of the user while walking. The GuideCane is considerably heavier than the white cane, because it uses a servo motor. The wheels are equipped with encoders to determine the relative motion. The servo motor, controlled by the built-in computer, can steer the wheels left and right relative to the cane. To detect obstacles, the GuideCane is equipped with ten ultrasonic sensors. A mini joystick located at the handle allows the user to specify a desired direction of motion. GuideCane is far heavier than the ordinary white cane and also it is hard to keep because it cannot be folded.

Smart Cane is one invention which was originally the creation of a common blind cane but it is equipped with a sensor system. This invention resembles GuideCane where this invention has a number of ultrasonic sensors and servo motors. This invention is designed with the aim at helping the blind in navigating. Ultrasonic sensors need to detect and avoid obstacles or objects located in front of the user. Meanwhile the fuzzy controller is required to determine the instructions that will be executed for example to turn right, left or stop. Like GuideCane, this invention also has a control button on the handle, and the button has four different directions. This invention has the same weaknesses as the GuideCane where there will be a problem to save space or to place the smart cane. Besides that, cost is also a weakness in this project as it uses ultrasonic sensors and a number of servo motors. If the cost is too high, users are not able to afford for it because the average income of the visually-impaired people is relatively small.

Smart Cane has been designed by students from Central Michigan University where this invention uses Radio Frequency Identification (RFID). RFID is used to detect objects or obstacles in front of the user and detects the RFID tag that has been placed in several areas to navigate the users. This invention is just like a normal stick but is equipped with a bag, worn by the user. The bag supplies electricity power to the invention and informs the user through speakers inside the bag. For users who do not have the ability to hear, there are special gloves that will vibrate at every finger, in which different vibrations in each finger have different meanings. However, this invention has several weaknesses and is only suitable for small areas. This is because it only detects the area with RFID tag otherwise this invention only works as a regular blind cane. In addition, this invention requires a high cost if it is used in the external environment because the larger area that need to be tagged, the higher cost is needed.

Mechatronic Blind Stick is a guiding system, designed to facilitate the daily work among the visually-impaired people. This invention has many similarities with the Smart Blind Cane. In which this invention uses ultrasonic sensors and sound vibrations. However, this invention also has several weaknesses; it cannot be folded and difficult to keep. In addition, this invention is not equipped with sensors to detect the water areas.

2.1 Software

MPLAB is software that is used to develop the source code of the PIC microcontroller. MPLAB is a Window ® based Integrated Development Environmental (IDE) for the Microchip Technology Incorporated PICmicro microcontroller families. It is allowed to write, debug and optimize the PICmicro applications' for firmware product design [13]. Besides that, this software includes a text editor, simulator, and project manager that makes programming becomes more schematic. MPLAB also support the MPLAB-ICE and PICMASTER ® emulators, PICSTART ® PLUS, and PROMATE ® II programmers [13]. Thus shows that MPLAB is compatible for various kinds of microchip development system tools. The reason of choosing MPLAB is because it is widely used and the language is easy to understand.

2.2 Hardware requirement

Ultrasonic sensors generate high frequency sound waves and evaluate the echo which is received back by the sensors. Sensors calculate the time interval between sending the signal and receiving the echo to determine the distance to an object. Ultrasonic is like an infrared where it will reflect on a surface in any shape. However, the ultrasonic has a better range detection compared to infrared. In robotic and automation industry, ultrasonic has been accepted well because of its usage [6, 9]. Magori and Walker [6] state that the endurance and accuracy of the sensor is not affected by physical contact. Comparing with other sensors, the ultrasonic is more accurate. Han and Hahn [13] have proven that the distance and angle measurements of ultrasonic are highly reliable by proving that the relative errors and variances of the measurements are within a reasonably small range [13]. These discussions explain that the ultrasonic is suitable for developing the Smart Blind Cane.

Microcontroller is a single chip that contains the processor (CPU), non-volatile memory for the program (ROM or flash), volatile memory for input and output (RAM), a clock and an I/O control unit and time [18]. It is designed for a small set of specific function to control a particular system. For example, microcontroller is used in wheelchair to controller the motion using remote control [8]. The reason of using microcontroller is because the microcontroller has the ability to store and run unique programs make it extremely versatile

A water detector is a small electronic device that is designed to detect the presence of water. According Hamelain [16], by using water sensor, as soon as it touches the water, it will short the circuit and this will cause a closed circuit then obtain the output that we desired. The water sensor is useful in a normally occupied area near any appliance that has the potential to leak water.

The aim of this paper is to discuss on a development work of an assistive tool for the visually-impaired people that alerts them of the obstacles in front, which is named SmartCane. This section elaborates the background foundations of the works in this study. Related works are discussed in supports of this study. Next, the steps in developing the assistive cane are addressed in detail. Further, the experiment including results and findings are elaborated at length. Finally, this paper concludes by discussing some possible works for the future.

## 3. Design and Development

This section describes the design and development process which outlined in Fig. 1.

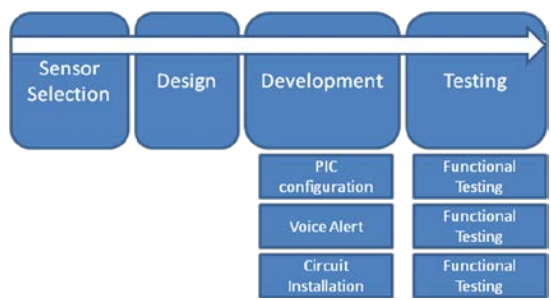

Fig. 1: Design and development process

### 3.1 Sensor Selection
This section describes the sensor selection process which ranging from information to the transmission frequency. Based on the following reasons, a 40 kHz transmission ultrasound signal is finally decided to be used.

   a) Size-40 kHz frequency which is produced by a 2-cm diameter transmission sensor. It can generate 2.4644 beams narrowness. This is an acceptable size to be installed in the blind cane.
   b) Interference-40 kHz is non-human audible frequency and that means it will not be affected by any noise that can be heard. Besides, noise generated by industries is normally ranging up to a few hundred kHz. Therefore this noise does not give effect to 40 kHz.
   c) Minimum attenuation-For an optimizing performance of ultrasonic sensor (to be able to detect range up to 1 meter), the attenuation of the transmitting signal and receiving signal is at minimum level. Since it free from the audible noises (<20 kHz) and industrial noises (MHz)

### 3.2 Design Process
The design process is based on the architecture. Basically the smart blind cane functions like ordinary blind canes. The difference is that the Smart Cane is equipped with ultrasonic sensor, water sensor and circuit box can be placed. Also the smart blind cane is designed to be foldable so that it is easy for the user to keep.

There are a few circuits designed. Firstly the design was tested on the proto-board for easy modification task. Once the circuit on the proto-board is finalized, printed circuit board (PCB) was fabricated. The important element when designing the PCB is the size of the PCB; which is long rather than wide. It is also to make sure that the design of Smart Cane is as small as possible.

After studying the overall process of the Smart Cane, the flowchart of the microcontroller was decided. Starting from the ultrasonic input, continue to the ADC process where the analogue data is converted to digital form. Then the output was generated. After the output was generated, the voice chip was reset. Next, the process continued to the water sensor input, where if water is detected, the buzzer is triggered. Then the buzzer is reset.

### 3.3 Development
There were two tasks involved in developing the Smart Cane. It started with code development, and physical installation.

*a) PIC Microcontroller Source Code Developments*

This involves creating the assembly language, based on prepared flow chart. Assembly (.asm) file is the source code written by the user. After creating the assembly file, MPLAB can transfer the assembly file into machine language i.e. the Hexa (.hex) file. In particular, Hexa file is the machine language to fuse the PIC microcontroller. Hexa file will not be able to produce if there are errors detected by the compiler in the assemble file. Fig. 2 shows a fraction of the source code for the ADC.

The source code in Fig. 2 explains when the sensor detected the object in closed range, the voice message is triggered, LED is on and the vibrator is on. When the sensor detected the object in medium range, the voice message is triggered, LED is on and the vibrator is on. Finally when the sensor detected the object in far range, the voice message is triggered, LED is on and the vibrator is off.

```
ADCON0=CHANNEL0;
read_adc();
```

```
        range=read_range();
        if(range<140)
          {
                led1=1;
                led2=0;
                led3=0;
                motor=1;
          }
        else if((range>140)&&(range<260))
          {
                led1=0;
                led2=1;
                led3=0;
                motor=1;
                delay(2000);
                motor=0;
                delay(2000);
                motor=1;
                delay(2000);
                motor=0;
                delay(2000);
                motor=1;
                delay(2000);
                motor=0;
                delay(2000);
          }
        else if((range>260)&&(range<400))
          {
                led1=0;
                led2=0;
                led3=1;
          }
        else        //
          {
                led1=0;
                led2=0;
                led3=0;
                motor=0;
          }
                delay(2000);
          }
        }
```

Fig. 2: The source code for ADC

b) *Voice Feedback*

As mentioned, the SmartCane can alert users about the distance in audio form. Hence, the voice circuit was set to enable the function, in which the design is depicted in Fig. 3.

As illustrated in Fig. 3, the ISD2560 is supplied with +5 V supply at pin 28 and 16 feet and pins 12 and 13 are connections to earth for the ISD2560 circuit. Meanwhile the pins 14 and 15 are connected to the loudspeaker or earphones (if not using speaker). The microphone is connected to pins 18 and 17 where both connections are connected in series with a 0.1uF capacitor. IN addition, pins 1 to 10 are inputs to ISD250 and the pins are connected to the PIC microcontroller. Besides, pins 22, 23, 24, 25, and 27 are also connected to the PIC microcontroller. Specifically, pin 23 (CE) is to enable the division operation, pin 24 (PD) is to put the device in standby mode, and pin 27 (P / R) functions to play back the recording. Meanwhile, pin 25 (eom) functions to put a throbbing low end of the order and pin 22 (OVF) is for a low throbbing in the end of memory array. On the PCB, the circuit is depicted in Figure 4.

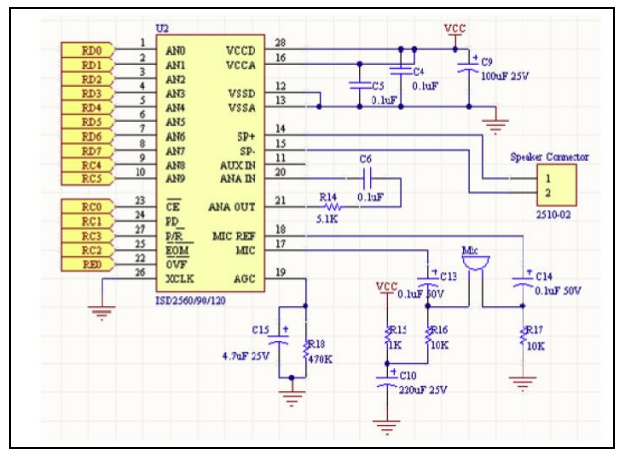

Fig. 3: Voice Chip Circuit

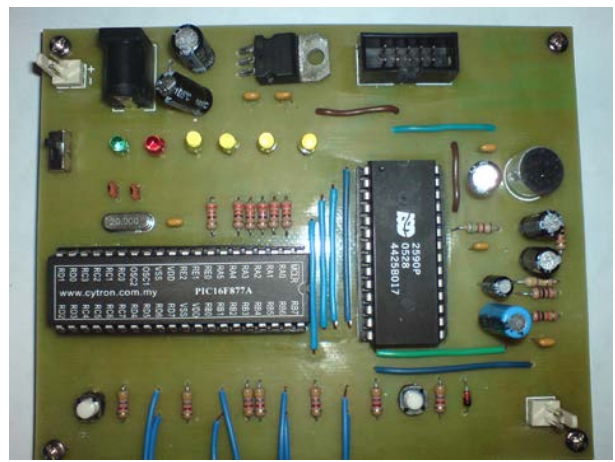

Fig. 4: The circuit printed in PCB

c) *Circuit Installation*

Since the size, cost, and appropriateness of the materials are well considered, the PVC is finally decided to be used. Holes in different sizes were drilled to house the sensors, power buttons and buzzer as the siren system. In order to avoid the impact of the shaking while using it, cotton was placed into the PVC to absorb the impact of shaking. Besides that, the circuits were covered by plastic also the connection between the jumpers was covered by sticker thoroughly. It is to avoid any short circuit occurrences of the circuit.

Fig. 5: PIC microcontroller circuit

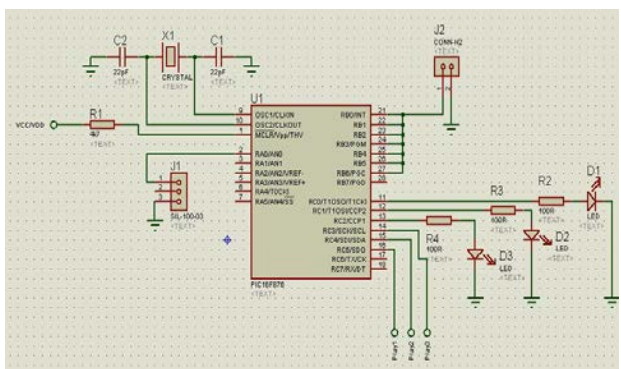

Fig. 5 depicts +5V power supply. The power supply is connected in series with R1 to pin 1 to always activate the PIC. R1 is used to restrict the flow through the PIC, while 8 MHz crystal oscillator and two 22pF capacitors connected to pins 9 and 10 to determine the project clock system. Analog inputs were connected on pin 2 meanwhile pin 3 to 7 are not used. As mentioned earlier, there are two types of output generated; the vibration and voice messages. Therefore, Pins 14, 15, and 16 were specified for connection to the voice chip message for voice output. Meanwhile pins 21 to 27 were connected in parallel where the pins that generate vibration. Analog inputs from ultrasonic sensors were connected to pin 2. IN addition, three LED were connected to pins 11, 12, and 13 on the PIC microcontroller where the LED works as a guide to the range object detection. IN particular, red LED is for the close range, meanwhile the yellow LED is for medium range and the green LED is for long. On the PCB, the circuit looks simple as shown in Figure 6.

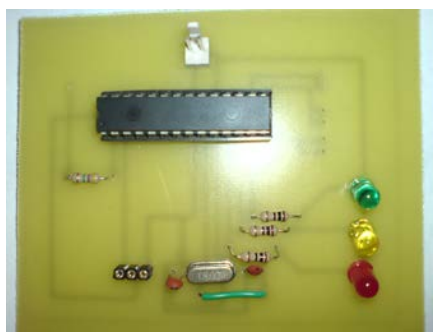

Fig. 6: The circuit printed in PCB

3.4 Testing

In this phase, the prototype was tested to make sure it functions as intended. Up to this point, this study has had opportunities to test the SmartCane in terms of its functionality. The results of the testing are discussed in the section 4.

## 4. Results and Analysis

This study has experimented on how well the SmartCane functions. To analyze that, this paper details the analysis into different modules. Fig. 7 visualizes users testing the SmartCane.

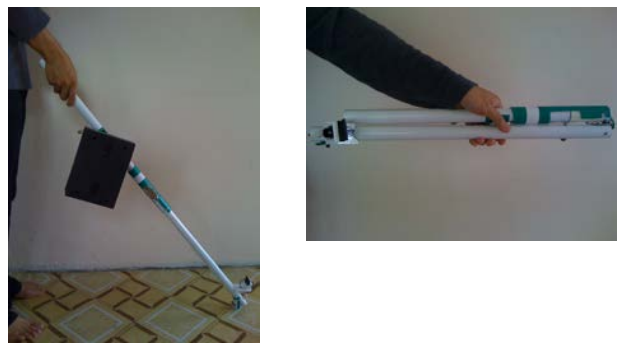

Fig. 7: Users testing the Smart Cane

### 4.1 Voice warning

Three different voice messages as listed in Table 1 are utilized to alert the user.

Table 1: Voice messages

| No | Voice Message | Distance |
|---|---|---|
| 1 | No object in 4 feet in front of you | Far |
| 2 | Objects are between 3 to 4 feet in front of you | Medium |
| 3 | An object is right in front of you | Close |

In the testing, the SmartCane was successfully addressing the three voice alerts. However, the first type of voice alert was found too misleading. Users feel confused with the voice alerts which are too repetitive.

### 4.2 Result of Ultrasonic Sensor Analysis

Besides the voice alert, this study also examines how the ultrasonic sensors function. The gathered data are presented in Table 2. The table shows analysis of the ultrasonic sensor analog voltage value between the calculation value and measurement value. Figure 8 illustrates the concepts of the calculation, which is followed with the formulas.

Table 2: Result of analyzing the ultrasonic sensor

| No | Range (Inches) | Calculation (mV) 1 inch = 10 mV | Measured (mV) | Error % ((cal)-(mea))/ cal * 100% |
|---|---|---|---|---|
| 1 | 0 | 0 | 0 | 0 |
| 2 | 5 | 50 | 42 | 16 |
| 3 | 10 | 100 | 94 | 6 |
| 4 | 15 | 150 | 142 | 5.3 |
| 5 | 20 | 200 | 193 | 3.5 |
| 6 | 25 | 250 | 245 | 2.0 |
| 7 | 30 | 300 | 296 | 1.3 |
| 8 | 35 | 350 | 346 | 1.1 |
| 9 | 40 | 400 | 396 | 1.0 |

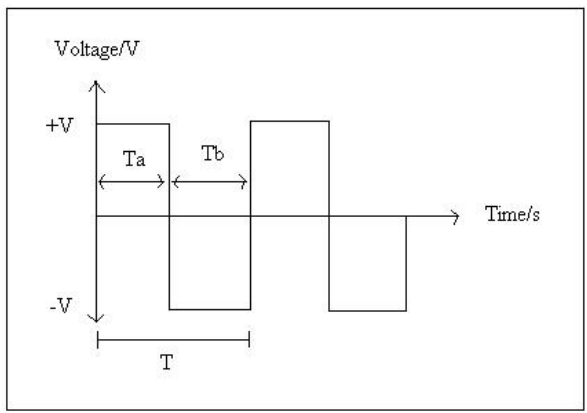

Fig. 8: Square Wave Oscillation

Figure 8 shows the usual square wave. In each time interval, the waves repeat regularly because they are stable. The waves commute between -V and +V, stop at every stage of running time (Ta or Tb). Therefore, T is the addition of Ta and Tb.

`T = Ta + Tb`

Where
  T is the period of the square wave
  Ta is the interval between $T_1$ and $T_2$
  Tb is the interval between $T_2$ and $T_3$
  For frequencies,

`F = (1 / T) Hz`

Where
  F is frequency
  T is period
  Therefore,
  `T = (1/F) = 0.69 C1[2R1 +(R2+R3)]`

By using the formula above, duty cycle can be found through the calculation steps as follows:
`T off = 0.69C1`
`T on = 0.69C1 (R1 +R2+R3)`
`Duty cycle = (T on / T)`
`Duty cycle = (R1+R2+R3) / [(R2+R3)+2R1]`

In Table 2, the voltage values obtained from the test are slightly different from the values shown in the ultrasonic sensor data sheet. It shows that there were errors with the ultrasonic analog output by. In relation, Figure 9 shows the difference between the calculation value and the measured value.

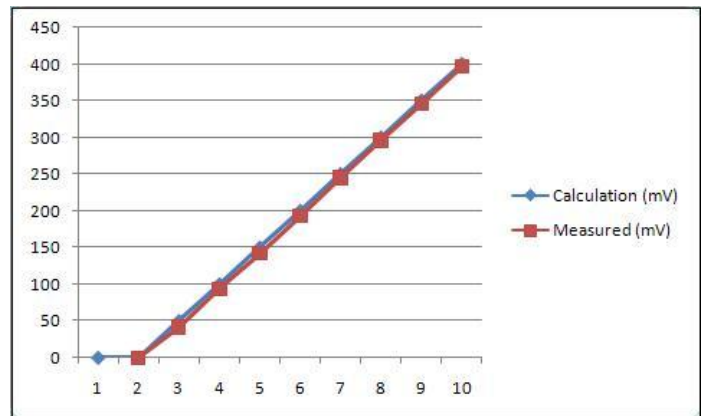

Fig. 9: Difference between the calculation value and the measured value

### 4.3 Result of analyzing the water sensor

There were few details that had been obtained when analyzing the water sensor as listed below:
1. The water sensor fully functions
2. The water sensor can detect if only the water is over 0.5 cm deep.
3. The water sensor buzzer cannot be stopped unless the water sensor is dry, so it needs to be wiped to stop the buzzer.

## 4. Conclusion and Recommendation

The main purpose of this study is to produce a prototype that can detect objects or obstacles in front of users and feeds warning back, in the forms of voice messages and vibration, to users. From the tests carried out on its functions reveal that the developed prototype which is named SmartCane has achieved its objectives.

This study would recommend that a power supply meter reading can be installed to monitor its power status. An alarm system also can be incorporated for use in a situation of very congested areas and replace PVC with steel so that it will be more durable and robust. In addition, a buzzer timer can be added so the buzzer will activate at a specific duration.